\documentclass[letterpaper,aps,prl,superscriptaddress,showpacs,floatfix,twocolumn]{revtex4}

\usepackage{graphicx}
\usepackage{hyperref}
\usepackage{amsmath}
\usepackage{url}
\usepackage{subfigure}

\usepackage{times}

\begin{document}

\title{Connected-Sea Partons }

\author{Keh-Fei Liu}
\affiliation{Department of Physics and Astronomy, University of
Kentucky, Lexington, Kentucky 40506, USA}

\author{Wen-Chen Chang}
\affiliation{Institute of Physics, Academia Sinica, Taipei 11529, Taiwan}

\author{Hai-Yang Cheng}
\affiliation{Institute of Physics, Academia Sinica, Taipei 11529, Taiwan}

\author{Jen-Chieh Peng}
\affiliation{Department of Physics, University of Illinois at
Urbana-Champaign, Urbana, Illinois 61801, USA}

\begin{abstract}
According to the path-integral formalism of the hadronic tensor, the nucleon sea contains
two distinct components called connected sea (CS) and disconnected
sea (DS). We discuss how the CS and DS are accessed in the lattice
QCD calculation of the moments of the parton distributions. We show
that the CS and DS components for $\bar u(x) + \bar d(x)$
can be extracted by using recent data on the strangeness parton
distribution, the CT10 global fit, and the lattice result of the
ratio of the strange to $u(d)$ moments in the
disconnected insertion. The extracted CS and DS for $\bar u(x) + \bar d(x)$
have distinct Bjorken $x$ dependence in qualitative agreement with
expectation. The analysis also shows that the momentum fraction of
the $\bar u(x) + \bar d(x)$ is about equally divided between CS and DS at
$Q^2 = 2.5 {\rm GeV}^2$.
Implications on future global analysis for parton distributions are presented.
\end{abstract}


\maketitle

There have been a number of developments in the understanding of the
flavor content of the nucleon sea, such as the observation of
the light-quark sea difference between $\bar d$ and $\bar u$ in Deep
Inelastic Scattering (DIS)~\cite{nmc91} and Drell-Yan processes~\cite{tma01},
the extraction of strange quark content $s + \bar s$ from semi-inclusive
DIS~\cite{hermes08}, and the lattice QCD calculations of sea quark
contributions to nucleon orbital angular momenta~\cite{ldd12}.
Evidence for the existence of intrinsic sea~\cite{bhp80} of the
light quarks has also been reported~\cite{wp11}.

Many theoretical models, including the meson cloud model, have been
suggested for describing the flavor structure of the nucleon sea~\cite{gp01}.
In order to gain new insights on the origins of the flavor content of the
nucleon sea, it is important to note that, according to the path-integral
formalism, there are two distinct sources for nucleon sea, namely, the
connected sea (CS) and the disconnected sea (DS)~\cite{ld95,liu00}.
The CS and DS are expected to have different shapes in their Bjorken-$x$
distribution, as well as distinct quark-flavor dependence. The first direct
experimental evidence for the existence of CS came from the observation of
large difference in the $\bar u(x)$ and $\bar d(x)$
distributions~\cite{ld95}. In this
paper we show that the two distinct contributions (CS and DS) to the
$\bar u(x) + \bar d(x)$ can be separated based on existing experimental data
and input from lattice QCD calculation.

The existence of the connected sea and disconnected sea can be illustrated
in the path-integral formalism of the hadronic tensor. In the Euclidean
path-integral formalism of the hadronic tensor $W_{\mu\nu}$, there are
three gauge invariant and topologically distinct diagrams, as shown in
Fig. 1. The various lines in Fig. 1 represent the quark propagators from
the source of the nucleon interpolation field at time $t=0$ to the sink time at $t$
and the currents are inserted at $t_1$ and $t_2$.

\begin{figure}[htbp] \label{hadonic_tensor}
\centering
\subfigure[]
{{\includegraphics[width=0.3\hsize]{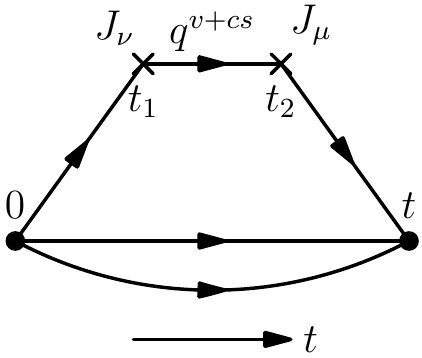}}
\label{val+CS}}
\subfigure[]
{{\includegraphics[width=0.3\hsize]{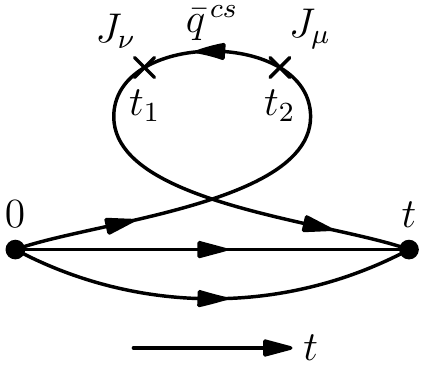}}
\label{CS}}
\subfigure[]
{{\includegraphics[width=0.3\hsize]{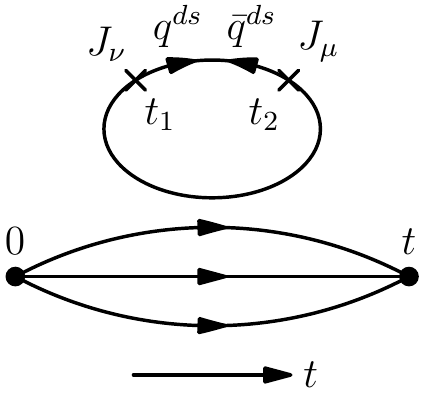}}
\label{DS}}
\caption{Three gauge invariant and topologically distinct diagrams
in the Euclidean path-integral
formalism of the nucleon hadronic tensor in the large momentum frame.
In between the currents
at $t_1$ and $t_2$, the parton degrees of freedom are
(a) the valence and CS partons $q^{v+cs}$, (b) the CS anti-partons
$\bar{q}^{cs}$, and (c) the DS partons $q^{ds}$ and
anti-partons $\bar{q}^{ds}$ with $q = u, d, s,$ and $c$. Only $u$
and $d$ are present in (a) and (b).}
\end{figure}

We first note that Fig.~\ref{CS}, where the quarks propagate backward in
time between $t_1$ and $t_2$, corresponds to the connected-sea anti-partons
$\bar{u}^{cs}$ and $\bar{d}^{cs}$. In contrast, the forward propagating
quarks in Fig.~\ref{val+CS} correspond to
valence and CS partons $u^{v + cs}$ and  $d^{v + cs}$,
where valence is defined as $q^v \equiv q^{v + cs} - \bar{q}^{cs}$
and $q^{cs}(x) \equiv \bar{q}^{cs}(x)$.
Finally, Fig.~\ref{DS} gives the DS $q^{ds}$ and $\bar{q}^{ds}$ for 
$q = u,d,s,c$, since it contains both forward and backward propagating quarks.
The nomenclature of connected and disconnected seas follows those in
the time-ordered perturbation theory -- CS is the higher Fock-state
component in the Z-graph where the quark lines are connected and the DS
corresponds to the vacuum polarization.

It is clear from Fig. 1 that the two sources of the sea quarks, CS and DS,
have interesting quark-flavor dependence. For example, while $u$ and
$d$ have both CS and DS, $s$ and $c$ have only DS. The small mass
difference between the $u$ and $d$ quarks implies that the DS cannot
account for the large $\bar d / \bar u$ difference observed in the DIS and
Drell-Yan experiments. Rather, this difference must originate primarily
from the CS diagram of Fig. 1(b) due to the fact that there are two 
$u$-valence quarks but only one $d$. The absence of the CS component for 
the strange and charm quarks also implies that any difference between 
$s(x)$ and $\bar s(x)$ (or $c(x)$ and $\bar c(x)$) distributions, as predicted 
in meson cloud~\cite{thomas} and intrinsic sea~\cite{brodsky} models, must come
from the DS diagram of Fig. 1 (c). The classification of parton distribution
functions in terms of flavor, CS and DS is given in Table~\ref{classification}.

\begin{figure}[htbp]
\centering
\subfigure[]
{\includegraphics[width=0.45\hsize]{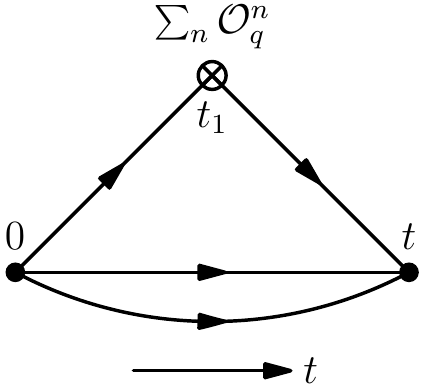}
\label{CI}}
\subfigure[]
{\includegraphics[width=0.45\hsize]{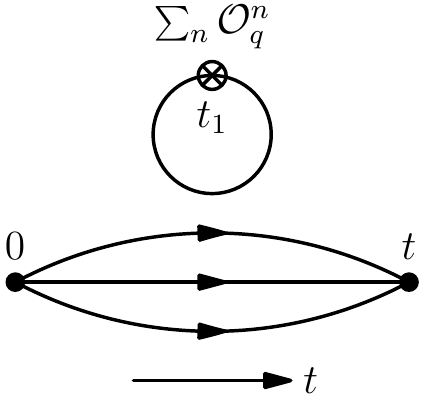}
\label{DI}}
\caption{The three-point functions after the short-distance expansion of
the hadronic tensor from Fig. 1.\,\, (a) The connected insertion (CI) is
derived from Fig.~\ref{val+CS} and Fig.~\ref{CS}. (b) The disconnected
insertion (DI) originates from Fig.~\ref{DS}. $\mathcal{O}_q^n$ are local
operators which are the same as derived from OPE.}
\label{fig:DI}
\end{figure}

%
\begin{table}[hbtp]
\caption{Classification of PDF in the nucleon for different flavors.}
\label{classification}
\vskip\baselineskip
\begin{center}
\begin{tabular}{|c|c|c|c|}
\hline
\multicolumn{4}{|c|}{Valence and Connected Sea}   \\
\hline
$u^{v+cs}(x)$ &  $\bar{u}^{cs}(x)$ & $d^{v+cs}(x)$ & $\bar{d}^{cs}(x)$ \\
\hline
\hline
\multicolumn{4}{|c|}{Disconnected Sea} \\
\hline
$ u^{ds}(x) + \bar{u}^{ds}(x)$   & $ d^{ds}(x) + \bar{d}^{ds}(x)$  & $ s(x)
+ \bar{s}(x)$ & $ c(x) + \bar{c}(x)$ \\
\hline
\end{tabular}
\end{center}
\end{table}

The CS and DS are also expected to have distinct distributions at the
small-$x$ region. Since there is only reggeon exchange for the flavor
non-singlet valence and CS, the valence and CS partons is 
$q^{v+cs}(x),\, \bar{q}^{cs}(x) {}_{\stackrel{\longrightarrow}{x \rightarrow 0}} 
\propto x^{-1/2}$ at small $x$.
For the DS partons, there is flavor-singlet pomeron exchange,
thus its small-$x$ behavior is $q^{ds}(x),\, \bar{q}^{ds}(x)
{}_{\stackrel{\longrightarrow}{x \rightarrow 0}} \propto x^{-1}.$
In addition, there are also meson cloud contributions which are prominent
in the medium-$x$ range. Partons with these different small-$x$ behaviors
are considered as extrinsic and intrinsic distributions for
charm~\cite{bhp80} and the light quarks~\cite{wp11}.
The distinct flavor and $x$-dependence of CS and
DS remain to be checked experimentally.

Under the short-distance expansion of the hadronic tensor between the current
insertions in the path-integral formalism ({\it N.B.} This corresponds
to operator product expansion (OPE) in the canonical formalism),
Fig.~\ref{val+CS} and Fig.~\ref{CS} become the connected insertions (CI) in
Fig.~\ref{CI} for a series of local operators $\sum_n O_q^n$ in the
three-point functions from which the nucleon matrix elements for the
moments of the CI are obtained. Here the flavor $q = u, d$ are the valence
flavors from the interpolation field. By the same token, the disconnected
four-point functions in Fig.~\ref{DS} become the disconnected insertions (DI)
in Fig.~\ref{DI} for the three-point functions to obtain the DI moments.
Here $q = u, d, s, c$ are the DS flavors in the DI.
The main advantage of the path-integral formalism over the canonical
formalism is that the parton degrees of freedom are tied to the topology
of the quark skeleton diagrams in Figs.~\ref{val+CS}, ~\ref{CS}, and~\ref{DS}
so that the CS and the DS can be separated.
Lattice QCD can access these three-point functions for the CI and DI which
separately contain the CS and DS and calculations of the moments of the
unpolarized and polarized PDFs for the quarks~\cite{lhpc10} and
glue~\cite{ldd12} have been carried out. Unfortunately, lattice
calculations cannot calculate the parton $x$-distributions
directly~\cite{liu00}, only moments are accessible.

To delineate the flavor and $x$ dependence of the parton distributions, it is
important to have the CS and DS separated in the global fitting as they
evolve differently in $Q^2$~\cite{liu00}. Once they are separated in one
$Q^2$, they will remain separated so that they can be used to fit experiments
or make predictions at other $Q^2$.
While the difference of $\bar{u}^{cs}(x)$ and $\bar{d}^{cs}(x)$ is
obtained from the E866 Drell-Yan~\cite{tma01} and HERMES
semi-inclusive~\cite{ack98} measurements of $\bar{d}(x) - \bar{u}(x)$,
there is not yet an established way to directly obtain
$\bar{u}^{cs}(x) + \bar{d}^{cs}(x)$ from experiments. We shall show how
to achieve this with a combination of experimental results and a lattice
calculation of $\langle x\rangle$ for the DI in Fig.~\ref{DI}.

\begin{figure}[htbp]
{\includegraphics[width=\hsize]{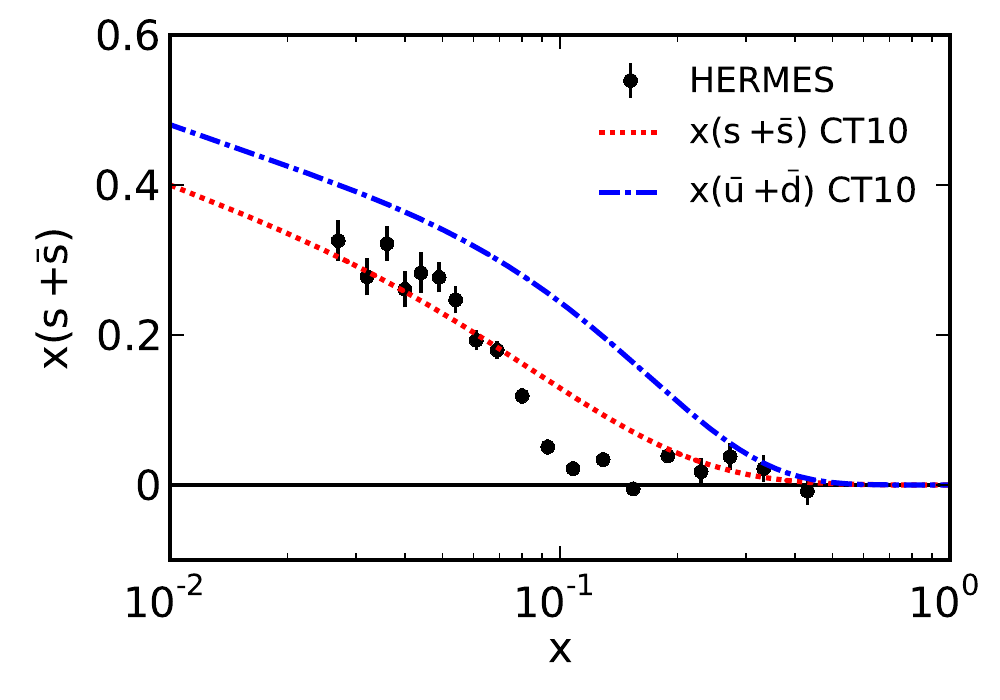}}
\caption{The strange quark PDF from HERMES semi-inclusive DIS experiment
of kaon production on deuteron. It is compared to CT10 results.}
\label{HERMES}
\end{figure}

Recent HERMES semi-inclusive DIS experiment of kaon production on
deuteron~\cite{hermes08} has produced the strangeness parton distribution
function $s(x) + \bar{s}(x)$ at $Q^2 = 2.5\,\, {\rm GeV}^2$ which is shown
in Fig.~\ref{HERMES}. One notable feature is that the values at medium $x$
around 0.1 are quite different from those obtained from the global fit of
CT10~\cite{CT10} (also drawn in Fig.~\ref{HERMES}), which did not include 
the HERMES data in the fit. Since $s$ and $\bar s$ are entirely
due to DS, the HERMES $s(x) + \bar s(x)$ data provide valuable information
on the shape of the $x$-distribution for DS, which is not available
from the lattice calculation.

We can now proceed to separate the CS and DS
components of the $\bar u + \bar d$ sea with the following approach.
First, we shall make the plausible ansatz that the distribution of
$\bar{u}^{ds}(x)+\bar{d}^{ds}(x)$ is proportional to that of
$s(x) + \bar{s}(x)$ and the proportionality is $\frac{1}{R}$, i.e.,
$\bar{u}^{ds}(x)+\bar{d}^{ds}(x) = \frac{1}{R}(s(x) + \bar{s}(x))$.
As discussed below, a recent lattice calculation has obtained
$R = 0.857 \pm 0.040$. We can then
extract $\bar{u}^{cs}(x)+\bar{d}^{cs}(x)$ from the difference of the CT10
result on $\bar{u}(x)+\bar{d}(x)$ and the HERMES data weighted with
$\frac{1}{R}$,
\begin{equation}  \label{udCS}
\bar{u}^{cs}(x)+\bar{d}^{cs}(x) = \bar{u}(x)+\bar{d}(x) - \frac{1}{R}(s(x)
+ \bar{s}(x))
\end{equation}
at $Q^2 = 2.5\,\, {\rm GeV}^2$. The $\bar u(x) + \bar d(x)$ in Eq. 1
can be taken from the recent CT10 PDF. 

A recent lattice calculation~\cite{doi08} of the momentum fraction
$\langle x\rangle$ for the strange and $u(d)$ in the DI was carried out
with $2+1$-flavor dynamical fermion gauge configurations
with the improved gauge and Wilson fermion (clover) actions. The ratio of
the momentum fraction between the strange and the $u(d)$ in the DI, where
multiplicative renormalization constants and some systematic errors cancel,
is extrapolated to the chiral limit with the lowest pion mass at
\mbox{600 MeV} and the result is
\begin{equation}  \label{ratio}
R=\frac{\langle x\rangle_{s+\bar{s}}}{\langle x\rangle_{u+\bar{u}}(DI)}
= \frac{\langle x\rangle_{s+\bar{s}}}{\langle x\rangle_{\bar{u}^{ds}+
\bar{d}^{ds}}} = 0.857(40).
\end{equation}
The second equality is based on the premises of isospin symmetry of the
DS, i.e., $\bar{u}^{ds} = \bar{d}^{ds}$ and that the parton-antiparton
difference is negligible in the DS, i.e., $u^{ds}=\bar{u}^{ds}$.

It is interesting to note that the ratio $R$ in Eq.~(\ref{ratio}) is close to unity,
much larger than the globally fitted ratio of
$\frac{\langle x\rangle_{s+\bar{s}}}{\langle x\rangle_{\bar{u}+\bar{d}}}
\sim 0.5$ as evidenced in the CT10 results in Fig.~\ref{HERMES}. The
difference is due to the fact that $\bar{u}+\bar{d}$ has an additional
contribution from CS which the strange partons do not have. Moreover,
the ratio $R = 0.857$ is very close to that of the CT10 fit at small
$x$ (e.g., $x < 2 \times 10^{-2}$). This is consistent with the recent
ATLAS measurement of the inclusive $W$ and $Z$ productions, where the
strange-to-down quark ratio was determined to be
$1.00 {}^{+0.25}_{-0.28}$ at $x=0.023$
and $Q^2 = 1.9\,\,{\rm GeV}^2$~\cite{ATLAS12}. Since partons at
small $x$ are dominated by the DS, this shows that
both the ratio of $(s(x)+\bar{s}(x))/(\bar{u}^{ds}(x)+\bar{d}^{ds}(x))$
of CT10 (and ATLAS) at small $x$ and the ratio of their second moments
from the lattice are practically the same. This supports the ansatz
that $\bar{u}^{ds}(x)+\bar{d}^{ds}(x) = \frac{1}{R}(s(x) + \bar{s}(x))$.

\begin{figure}[htbp]
\centering
{ {\includegraphics[width=\hsize]{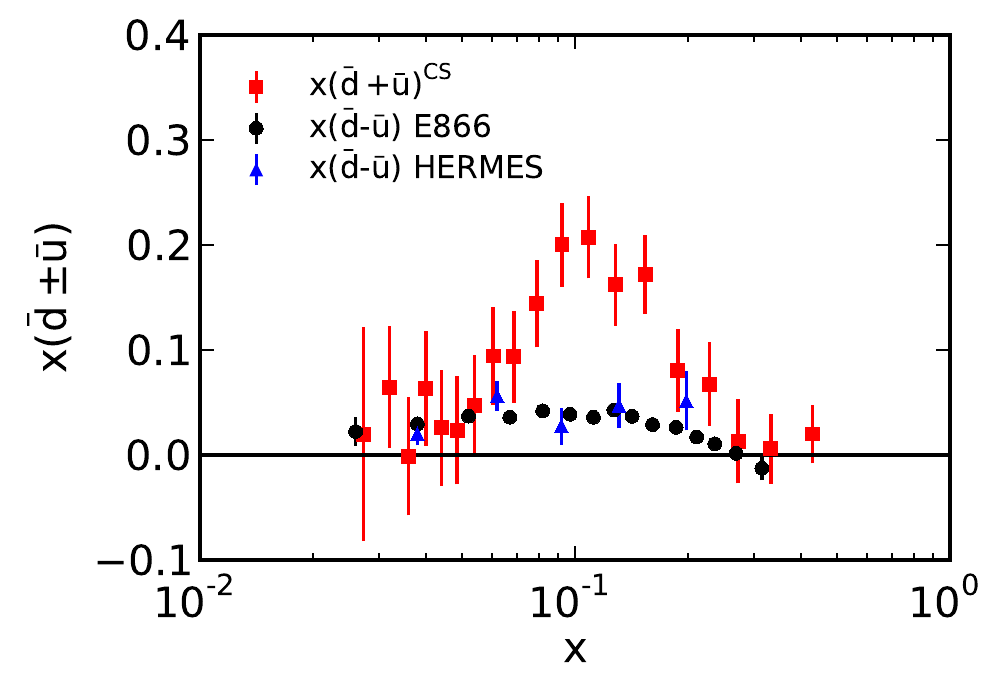}\ \ \ } }
\caption{$x(\bar{d}^{cs}(x) + \bar{u}^{cs}(x))$ obtained
from \protect{Eq.~(\ref{udCS})} is plotted together with
$x(\bar{d}(x) - \bar{u}(x))$ from E866 Drell-Yan
experiment~\protect{\cite{tma01}} and from SIDIS HERMES
experiment~\protect{\cite{ack98}}.}
\label{CSupmd}
\end{figure}

We plot the distribution function evaluated with Eq.~(\ref{udCS}), multiplied by the
momentum fraction, i.e., $x(\bar{u}(x)+\bar{d}(x) - \frac{1}{R}(s(x)
+ \bar{s}(x))$ in Fig.~\ref{CSupmd} together with $x(\bar{d}(x)-\bar{u}(x))$
from E866 Drell-Yan measurement~\cite{tma01}
at $Q^2 = 54\,\,{\rm GeV}^2$ and from SIDIS HERMES measurement~\cite{ack98} at
$\langle Q^2\rangle = 2.3\,\,{\rm GeV}^2$.
We see that $x(\bar{u}^{cs}(x)+\bar{d}^{cs}(x))$ from Eq.~(\ref{udCS})
is peaked at medium $x \sim 0.1$, the same way as
$x(\bar{d}(x) - \bar{u}(x))$ from E866 and \mbox{HERMES}. This is consistent
with the expectation that the small-$x$
of CS, like the valence, behaves as $x^{-1/2}$ as we alluded to earlier;
so that, when CS is multiplied with $x$, it would be peaked at medium $x$,
in contrast to that of the DS, e.g., $x(s(x)+\bar{s}(s))$ in Fig.~\ref{HERMES}.
Furthermore, we note that $x(\bar{u}^{cs}(x)+\bar{d}^{cs}(x))$ is generally
larger than $x(\bar{d}(x) - \bar{u}(x))$ in this $x$-range as it should
and is larger by a factor $\sim$ 4 at the peak.

\begin{figure}[hbtp]
\centering
{ {\includegraphics[width=0.9\hsize]{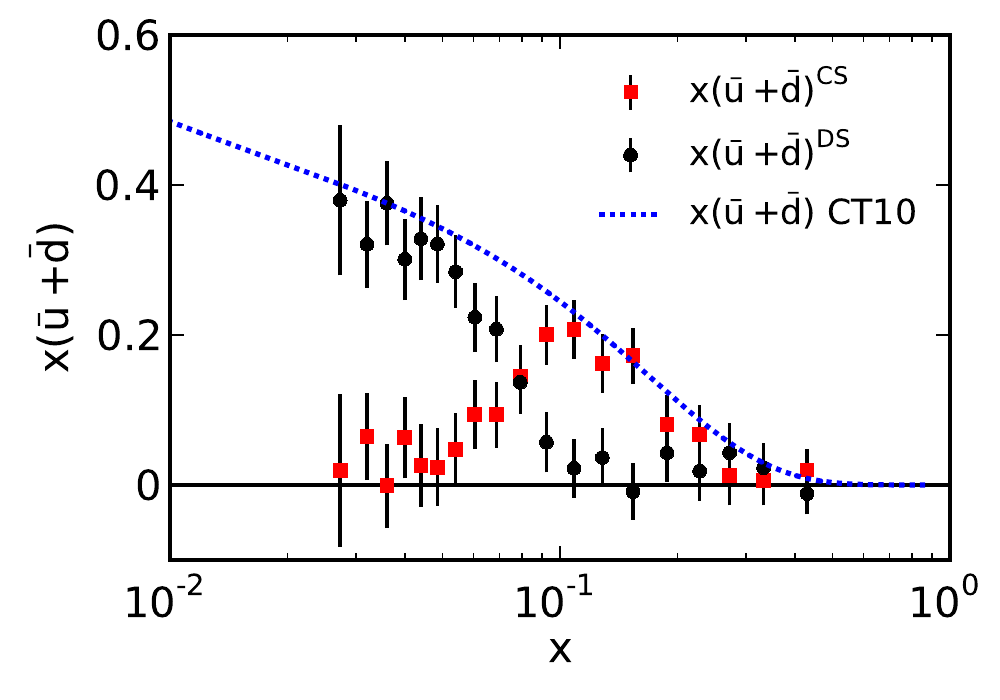}\ \ \ } }
\caption{$x(\bar{u}^{cs}(x) + \bar{d}^{cs}(x))$ obtained
from \protect{Eq.~(\ref{udCS})} is plotted together with
$x(\bar{u}(x) + \bar{d}(x))$ from CT10 and $\frac{1}{R}x(s(x)+\bar{s}(x))$
which is taken to be $x(\bar{u}^{ds}(x)+\bar{d}^{ds}(x))$.}
\label{CSu+d}
\end{figure}

We also plot $x(\bar{u}(x)+\bar{d}(x) - \frac{1}{R}(s(x) + \bar{s}(x))$,
$x(\bar{u}^{ds}(x)+\bar{d}^{ds}(x) = \frac{1}{R}x(s(x) + \bar{s}(x))$
and $x(\bar{u}(x)+\bar{d}(x))$ from CT10 in Fig.~\ref{CSu+d} to show
that the CS and DS have very different $x$-dependence. The different
shapes of CS and DS are in good agreement with the
expectation discussed earlier. This agreement lends support for the
approach we adopted. It is interesting to note that should a very
different value of $R$ be used, the $x$-dependence of CS and DS would
no longer agree with expectation. In particular, if $R$ were appreciably 
larger than the present value of $0.857$, Eq.~(\ref{udCS}) would lead to 
a CS whose small $x$ behavior would be $\sim x^{-1}$ which is inconsistent 
with the fact that CS is from the connected 
insertion. In this case, $x(\bar{u}^{cs}(x)+\bar{d}^{cs}(x))$ would approach 
a constant as $x \rightarrow 0$ as opposed to zero, as illustrated in 
Fig.~\ref{CSu+d}. On the other hand, if $R$ were appreciably smaller 
than $0.857$, the CS from Eq.~(\ref{udCS}) would turn out to be negative 
and this would not be commensurate with the probability interpretation for CS.

We can also calculate the momentum fractions carried by the
CS and DS of $\bar{u} + \bar{d}$ as follows. First,
the total contributions of $\bar{u} + \bar{d}$, i.e.,
\mbox{$\langle x\rangle_{\bar{d}+\bar{u}} = \int_0^1 dx\, x
(\bar{d}(x)+\bar{u}(x))$} for the global fittings
of CT10, CTEQ6~\cite{CTEQ6} and MSTW08~\cite{MSTW08} are listed in
Table II. Here, the uncertainty of a specific PDF function is estimated
by varying the range of the calculated quantity
with input from the alternative eigenvectors of the PDF and is
quoted in the parentheses. The CS contributions
$\langle x\rangle_{\bar{d}^{cs}+\bar{u}^{cs}}$
in Table~\ref{momentum_fraction} are then obtained from
Eq.~\ref{udCS} with inputs of $\bar{d}(x)+\bar{u}(x)$ from the different
PDFs and integrated over the range $0.025 < x < 0.48$ plus
a small contribution outside this range which is estimated to be 2\% from
the fractional contribution to $x(\bar{d}(x) - \bar{u}(x))$ outside this
range based on CT10. The total uncertainty includes the contributions
from HERMES data, the lattice calculation of $R$, the PDF, and the
unmeasured $x$ region. The corresponding
$\langle x\rangle_{\bar{d}^{ds}+\bar{u}^{ds}}$ in
Table~\ref{momentum_fraction}
is taken to be the difference between the respective
$\langle x\rangle_{\bar{d}+\bar{u}}$ and
$\langle x\rangle_{\bar{d}^{cs}+\bar{u}^{cs}}$. Finally, we give the
ratio $\frac{\langle x\rangle_{\bar{d}^{cs}+\bar{u}^{cs}}}{\langle
x\rangle_{\bar{d}^{ds}+\bar{u}^{ds}}}$. It is interesting that this
ratio is close to unity, showing that the momentum fraction of
$\bar{d} + \bar{u}$ is about equally divided between the CS and the DS
at this low $Q^2$. Future lattice calculations could provide a direct check of the
second moment of $\bar{u}^{ds}(x)+\bar{d}^{ds}(x)$ shown in Table II.

\begin{table}[tbp]   
\caption {Values of various moments using the HERMES data,
the lattice QCD result, and three different PDFs at
$Q^2=2.5\,\,{\rm GeV}^2$.}
\label{momentum_fraction}
\begin{center}
\begin{tabular}{|c|c|c|c|}
\hline
 & CT10 & CTEQ6 & MSTW08 \\
\hline
 $\langle x \rangle _{\bar{d} + \bar{u}}$ & 0.0639(14) & 0.0614(14) & 0.0690(11) \\
 \hline
 $\langle x \rangle_{\bar{d}^{cs} + \bar{u}^{cs}} $ & $0.0294(54)$ & $0.0281(54)$ &
$0.0347(53)$ \\
  \hline
 $\langle x \rangle_{\bar{d}^{ds} + \bar{u}^{ds}}$ & $0.0344(52)$ & $0.0332(52)$ & $0.0342(52)$\\
 \hline
 $\frac{\langle x \rangle_{\bar{d}^{cs} + \bar{u}^{cs}}}{\langle x \rangle_{\bar{d}^{ds} + \bar{u}^{ds}}}$
  & $0.86(29)$ & $0.85(30)$ & $1.02(32)$ \\
\hline
\end{tabular}
\end{center}
\end{table}

In order to gain deeper and more precise understanding of the PDF in terms
of their flavor, $x$ and $Q^2$ dependence, it is essential to have the CS
and DS separately accommodated in the extended evolution equations. Only
then will they remain separated at different $Q^2$ to facilitate global
fitting, as they evolve differently with the CS evolving like the
valence~\cite{liu00}. This will have an impact on the
gluon distribution as well. Furthermore, only with CS and DS separated,
will one be able to address the flavor dependence, i.e.,
$\bar{u}^{ds} \neq \bar{d}^{ds}$ and parton-antiparton difference of
the $u$ and $d$ partons in the DS and check the validity of the ansatz
that $\bar{u}^{ds}(x) + \bar{d}^{ds}(x)$ is
proportional to $s(x) + \bar{s}(x)$. As the lattice calculations are
getting more refined when the physical pion mass, continuum limit, and
large volume limit are approached, they could serve as valuable constraints
for the parton moments. In particular, higher moments in CI will help
separate the valence and CS parton distributions in the global analysis
and the DI calculation of the fourth moment for the strange and $u/d$
(i.e., $\langle x^3 \rangle_u (DI) = \int dx\, x^3(u^{ds}(x) +
\bar{u}^{ds}(x))$) can be used to gauge how good the proportionality
assumption is about their distributions.

In summary, we have shown that there are two sources for
the sea partons, the CS and DS, based on the path-integral
formalism of the hadronic tensor. While the $u$ and $d$ have
both CS and DS contributions, the $s$ and $c$ partons are from
the DS only. We also expect that the CS and DS have different
$x$ distributions. These different flavor and $x$ dependence
offer the possibility of disentangling the CS from the DS. We first
show that the expectation for the dominance of DS at small $x$ is
supported by the good agreement between the lattice calculation of
$R=\frac{\langle x\rangle_{s+\bar{s}}}{\langle
x\rangle_{u+\bar{u}}(DI)}=0.857(40)$ and the
ratio of $x(s(x)+\bar{s}(x))$ from the HERMES data~\cite{hermes08} to
$x(\bar{u}(x)+\bar{d}(x))$ from CT10 at small $x$ (e.g., $x < 2 \times 10^{-2}$)
as shown in Fig~\ref{HERMES}. 
Given this agreement, we show how the \mbox{HERMES} data on strangeness
parton distributions, the lattice calculation of $R$, and the CT10
global fit of $\bar{d}+\bar{u}$, can determine the separate CS and DS contributions 
in $\bar{u}+\bar{d}$.
We stress that the ansatz of the proportionality between
$\bar{u}^{ds}(x)+\bar{d}^{ds}(x)$ and $s(x)+\bar{s}(x)$ should be checked
with lattice calculation of $R$ with light dynamical fermions as well as the ratio for
$\langle x^3\rangle$ in the DI. Future global analysis of PDF should
have CS and DS separated in the fitting and in the evolution equations.

This work is partially supported by U.S. Department of Energy grant DE-FG05-84ER40154, 
National Science Foundation and the National Science Council of the Republic of China. 
We thank C.P. Yuan for fruitful discussion.

\end{document}